# Quantum Information through Angular momentum of Photon

Dipti Banerjee# and Dipan Sinha*
Department of Physics
Vidyasagar College for Women,
39, Sankar Ghosh Lane,
Kolkata-700006

**Abstract**
The angular momentum of photons is the key source of quantum information. The transfer angular momentum is possible as circularly polarized (CP) light passed through wave plates. The twisted birefringent medium behaves as 'Q' plate. The passage of CP through two consecutive half wave plates traces a closed curve on Poincare sphere. As a result the geometric phase in association with gain of orbital angular momentum is developed.

Light can carry angular momentum in two ways: spin angular momentum (**SAM**) and orbital angular momentum (**OAM**). In **circularly polarized light** every polarization vector rotates and the light has *spin angular momentum* (SAM)= $\pm h/2\pi$. The sign of the **SAM** depends on whether the light is left- or right circularly polarized. It may be noted that photons in a **linearly polarized** light beam carry no **SAM** [1]. Apart from the **SAM**, photons can also carry orbital angular momentum **OAM** arising from the inclination of the phase fronts with respect to beam's propagation axis. These beams have an azimuthal phase dependence $exp(im\varphi)$, where $\varphi$ is the azimuthal angle and $m$ is the OAM index. The value of m is an integer and corresponds to the number of times the phase changes by $2\pi$ in a closed loop around the beam. The beam has an angular momentum if it possesses an azimuthal component the angular momentum density $j_z$ along the propagation direction z [2]

$$j_z = \varepsilon_0 \omega \left( l|u|^2 - \frac{1}{2}\sigma r \frac{\partial |u|^2}{\partial r} \right)$$

$$J = l\hbar + \sigma\hbar \qquad (1)$$

The phase fronts of light beams in orbital angular momentum (**OAM**) eigenstates rotate clockwise for positive **OAM** values, anticlockwise for negative values. The phase front with 0 **OAM** doesn't rotate at all.
The **SAM** space is a conventional Poincare sphere where with the change of polarization from point to point three kinds of polarized states (circular, elliptical and linearly) are defined. For every **OAM**, there are two ±1 polarized photon. There exists 2 to 1 correspondence between **SAM** and **OAM** sphere.
 As m can take any integer, there could be large number of possible eigenstates of **OAM**.
 In principle, a single photon can in this way carry an arbitrarily large amount of information. The only trouble was that no method had been found that could distinguish all these OAM states with good efficiency.

#email: deepbancu@hotmail.com
*Ph.D.Scholar of CU

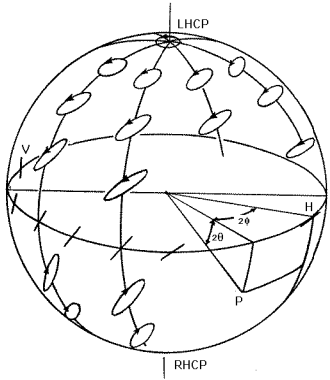 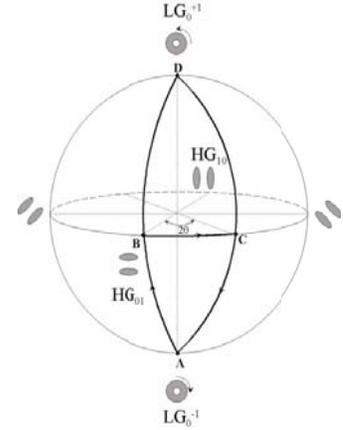

Fig1: SAM (Poincare) Sphere  Fig2: OAM Sphere

The **OAM** and **SAM** spheres are analogous. Addition of RCP & LCP produce linear states similar as $LG_0^{\ 1} + LG_0^{-1}$ to produce $HG_{1,0}$ [3]. Photon **SAM** can be manipulated by polarizers & birefringent plates. Cylindrical lens converters & Dove prisms can be used for changing **OAM** slowly. **OAM** light beam has increased interest due to wide range of scientific and technological applications both in classical and quantum regimes of light. Karimi, Santamato [4] pointed out that **OAM** can be further divided into two components:
1. External **OAM** – Cross product of the total momentum transported by the beam and the position of its axis relative to the origin of coordinates.
2. Internal **OAM** -- Associated with the helical structure of the optical wave-front around the beam axis. It acts as an additional spin of the whole beam around the axis.

Spin to orbital conversion of angular momentum is very important where variation of SAM occurring from medium's birefringence gives rise to the appearance of OAM, arising from medium's in-homogeneity. In quantum information field the interest of OAM degree of freedom of light mainly arises from the possibility of using its higher-dimensionality for encoding a large amount of information in single photons.

The standard unit of quantum information is a qubit, the microscopic system such as an atom, a nuclear spin or a polarized photon that can exist in an arbitrary superposition α|0>+β|1> as a computational resource.

Using the elementary qubits $|0\rangle = \begin{pmatrix} 1 \\ 0 \end{pmatrix}; |1\rangle = \begin{pmatrix} 0 \\ 1 \end{pmatrix}$ and quantum gates any single qubit can be prepared.

$$|0\rangle \text{---------[H]------}2\theta\text{------[H]-----}\pi/2+\theta = \cos\theta|0\rangle + e^{i\phi}\sin\theta|1\rangle \qquad (2)$$

Photon is a two state system. To describe polarization state in the orthogonal basis formed by the linear polarization states in the horizontal and vertical direction

Circularly polarized states are $|L\rangle = \frac{1}{\sqrt{2}}\begin{pmatrix} 1 \\ i \end{pmatrix}, |R\rangle = \frac{1}{\sqrt{2}}\begin{pmatrix} i \\ 1 \end{pmatrix}$

A general photon state can be written as $|\psi\rangle = C_\updownarrow|0\rangle + C_\leftrightarrow|1\rangle = \begin{pmatrix} C_\updownarrow \\ C_\leftrightarrow \end{pmatrix}$

Quantum theory describes a physical system in terms of state vectors in a linear space. If two systems are combined then they have correlations between observations and the state is called entangled. An entangled state between a pair of two-level systems is called a **singlet** state if it is of the form

$$|\Psi_{\sin glet}\rangle = \frac{1}{\sqrt{2}}(|\uparrow\rangle_1|\downarrow\rangle_2 - |\downarrow\rangle_1|\uparrow\rangle_2) = \frac{1}{\sqrt{2}}(|\uparrow\rangle_1 \quad |\downarrow\rangle_1)\begin{bmatrix} 0 & 1 \\ -1 & 0 \end{bmatrix}\begin{pmatrix} |\uparrow\rangle_2 \\ |\downarrow\rangle_2 \end{pmatrix}$$

As the spin vector is rotated by an arbitrary angle Θ, the transformation results the correlation un-affected and giving rise another singlet state [5].

$$\begin{pmatrix}|+\rangle\\|-\rangle\end{pmatrix}=\begin{pmatrix}\cos\theta & \sin\theta\\-\sin\theta & \cos\theta\end{pmatrix}\begin{pmatrix}|\uparrow\rangle\\|\downarrow\rangle\end{pmatrix}; |\Psi_{singlet}\rangle=\frac{1}{\sqrt{2}}(|+\rangle_1|-\rangle_2-|-\rangle_1|+\rangle_2) \quad (3)$$

Thus whichever spin direction is chosen for the observation, the values found in systems are opposite. This property will prevail however large the separation between the two systems and is also independent of the order of the measurements. The term **entanglement** thus refers to a certain division of total space. In this article we like to mention that OAM can be readily combined with other degrees of freedom of the photon. Q-plate is a device [4] which is an interesting potential application of combining SAM and OAM to create a frame –invariant encoding of quantum information. It is a birefringent plate made of a vortex-patterned liquid crystal film can imprint its topological charge into the optical phase of the incident beam. The plate affects the Photons **SAM** too. Easy control of **OAM** could be possible by either changing its polarization or changing retardation of the plate. If the birefringent plate is q, the OAM of light beam passing through such a 'q-plate changes by an amount ±2qℏ per photon.

The action of q-plate on a single photon quantum state [4] can be represented considering the initial state as $|P,m\rangle=|P\rangle_\pi|m\rangle_0$

Where P stands for the polarization state and m is the OAM value in units of ℏ. The q-plate action which can be described by the following rules:

$$\overline{QP}|L\rangle_\pi|m\rangle_0=|R\rangle_\pi|m+2q\rangle_0 \quad (4)$$
$$\overline{QP}|R\rangle_\pi|m\rangle_0=|L\rangle_\pi|m-2q\rangle_0$$

If the applied input is a linearly polarized light having m=0, we obtain the following output state:

$$\overline{QP}|H\rangle_\pi|0\rangle_0=\frac{1}{\sqrt{2}}(|L\rangle_\pi|-2q\rangle_0+|R\rangle_\pi|2q\rangle_o) \quad (5)$$

This shows that output polarized light can be interpreted as an entangled state of the SAM and OAM degrees of freedom of the same photon. Its action introduces a controlled coupling between polarization (SAM) and OAM that can be conveniently exploited in many different ways.

Inspired by this work we here shall highlight the effect of polarized light on twisted wave plates which is the birefringent plate of specific thickness.

Considering the incident polarized light as $|\psi\rangle=\begin{pmatrix}\cos\frac{\theta}{2}e^{i\phi}\\\sin\frac{\theta}{2}\end{pmatrix}$

Following Jones [6] and Berry [7] a polarization matrix M=(|ψ⟩⟨ψ|-1/2) can be determined as

$$M=1/2\begin{pmatrix}\cos\theta & \sin\theta e^{i\phi}\\\sin\theta e^{-i\phi} & -\cos\theta\end{pmatrix}=1/2\begin{pmatrix}Y_1^0 & Y_1^1\\Y_1^{-1} & Y_1^0\end{pmatrix} \quad (6)$$

It has been pointed out [8] that this polarization matrix for this case is parameterized by Θ,φ lies on the sphere as equivalent as OAM sphere for \$l=1. For higher OAM states l=2,3.., further study is needed to evaluate the polarization matrix for a particular OAM from the respective product harmonics $Y_l^m$. SAM of polarized photon is associated with optical polarization and as consequence the parameter in connection with **helicity** changes. To represent SAM/ helicity the parameter χ is introduced that extend the Poincare sphere whose picture is seen in the following fig. SAM space is possible to realize by parameters Θ,χ.

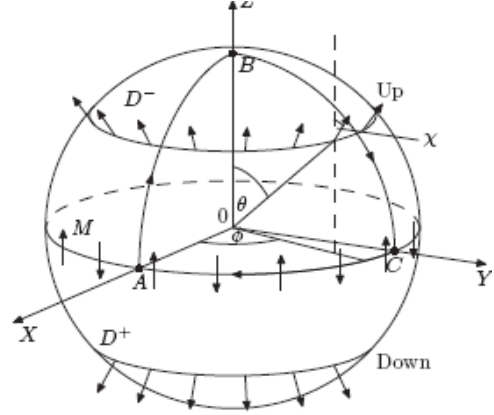

**Figure 1.** Geometrical interpretation of the quantized spinor.

Since for every OAM, there are two polarized photon for right handed state (spin parallel to motion) and left handed state (spin opposite to motion), it is very obvious there exists 2 to 1 correspondence between SAM and OAM space. We may realize for every OAM sphere there exist two SAM hemispheres.

The property of birefringence of the optical medium can be represented by the differential matrix N. At a particular position z of optical medium, Jones showed that the spatial variation of the polarization matrix M develop the N matrix,

$$N = \frac{dM}{dz}M^{-1} = \frac{dM}{d\theta}\frac{d\theta}{dz}M^{-1} \tag{7}$$

Considering z=cosΘ, the thickness of the optical medium, the N matrix can be obtained from M of eq.(6), where Θ is the angular variable of light after refraction.

$$N = \eta \begin{pmatrix} 0 & -e^{i\phi} \\ e^{-i\phi} & 0 \end{pmatrix} \tag{8}$$

The internal birefringence η [9] visualized through eigen values ±iη have dependence on Θ. As the homogeneous birefringent crystal is uniformly twisted about the direction of transmission, the N matrices are transformed upon rotation. When the angle of twist Θ is independent of crystal thickness, the twisted birefringent medium becomes

$$N' = S(\theta)NS(-\theta) \tag{9}$$

It takes the form

$$N' = \eta\cos\Phi \begin{pmatrix} 0 & -1 \\ 1 & 0 \end{pmatrix} - i\eta \sin\Phi \begin{pmatrix} -\sin 2\theta & \cos 2\theta \\ \cos 2\theta & \sin 2\theta \end{pmatrix} \tag{10}$$

For a half wave- plate, Φ=π/2 in N

Thus $N\left(\frac{\pi}{2}\right) = H = -i\eta \begin{pmatrix} 0 & 1 \\ 1 & 0 \end{pmatrix}$. The passage of LCP through half wave plate changes the SAM resulting it to $H|L> = -i\eta|R>$

Now twist by an angle θ results, $N_\theta\left(\frac{\pi}{2}\right) = S(\theta)HS(\theta)^{-1} = H' = -i\eta \begin{pmatrix} -\sin 2\theta & \cos 2\theta \\ \cos 2\theta & \sin 2\theta \end{pmatrix} \tag{11}$

The passage of left circularly polarized light through twisted optical half wave plate results

$$H'|L> = -i\eta \begin{pmatrix} -\sin 2\theta & \cos 2\theta \\ \cos 2\theta & \sin 2\theta \end{pmatrix}\frac{1}{\sqrt{2}}\begin{pmatrix} 1 \\ i \end{pmatrix} = -\frac{i\eta}{\sqrt{2}}\begin{pmatrix} -\sin 2\theta + i\cos 2\theta \\ \cos 2\theta + i\sin 2\theta \end{pmatrix} = -\frac{i\eta}{\sqrt{2}}\begin{pmatrix} ie^{i2\theta} \\ e^{i2\theta} \end{pmatrix} \tag{12}$$

$H'|L> = -\frac{i\eta}{\sqrt{2}}|R> e^{i2\theta}$

This shows twisted half wave-plate change the SAM ( $|L>$ to $|R>$) and OAM of the incident polarized light. Twisted half wave-plate change both the SAM ( $|L>$ to $|R>$) and OAM of the incident polarized light. Since a linearly polarized light is a combination of both LCP and RCP, $|H>=|L>+|R>$, the passage of this state through H' shows

$$H'|H> = \eta(|R> e^{i2\theta} + |L> e^{-i2\theta})  \quad (13)$$

the entangled state of SAM and OAM state similar as in eq.5 in connection with Q plate.

If the polarized light from first wave plate H is made to pass through another wave plate H' the combination is as similar [10] as.

$$HH' = -\eta^2 \begin{pmatrix} 0 & 1 \\ 1 & 0 \end{pmatrix} \begin{pmatrix} -\sin 2\theta & \cos 2\theta \\ \cos 2\theta & \sin 2\theta \end{pmatrix} \quad (14)$$

$$= -\eta^2 \begin{pmatrix} \cos 2\theta & \sin 2\theta \\ -\sin 2\theta & \cos 2\theta \end{pmatrix}$$

Here as the LCP is made to pass through two consecutive half wave plates one is twisted from other by an angle θ

$$HH'|L> = -\frac{\eta^2}{\sqrt{2}} \begin{pmatrix} \cos 2\theta & \sin 2\theta \\ -\sin 2\theta & \cos 2\theta \end{pmatrix} \begin{pmatrix} 1 \\ i \end{pmatrix} = -\frac{\eta^2}{\sqrt{2}} \begin{pmatrix} \cos 2\theta + i \sin 2\theta \\ -\sin 2\theta + i \cos 2\theta \end{pmatrix}$$

$$= -\frac{\eta^2}{\sqrt{2}} \begin{pmatrix} 1 \\ i \end{pmatrix} e^{i2\theta} = -\eta^2 |L> e^{i2\theta} \quad (15)$$

In analogy with eq.4, as incident LCP is rotated by an arbitrary angle θ, the transformation results the correlation un-affected. The polarization of the incident light (SAM) passing through two consecutive wave plates is preserved. The arrangement of plates are such that the incident polarized light could trace a closed area on Poincare sphere. The effect of twist helps to achieve the geometric phase of Pancharatnam [11] as the initial state $|A>$ of polarized light unite with the final $|A'>$ after a rotation.

$<A|A'> = \exp(i\gamma(C)/2)$

where γ(C) is the solid angle swept out by the area enclosed on its unit sphere. The geometric phase in connection with OAM is acquired.

So $\langle L|HH'|L\rangle = -\eta^2 e^{i2\theta}$ (16)

and $\langle R|HH'|L\rangle = 0$

Similarly $\langle R|HH'|R\rangle = -\eta^2 e^{-i2\theta}$

Due to the orthogonal properties of $|L>$ and $|R>$, the linearly polarized light visualize geometric

phase $<H|HH'|H> = <L|HH'|L> + <R|HH'|R> = -\frac{\eta^2}{2} \cos 2\theta$ (17)

after passing through two half wave plates.

If we apply arbitrary twist of angle α between the two H plates, LCP acquire similar OAM dependent geometric phase in terms of 2α without dependence on SAM.

$H(\phi+\alpha)H(\phi)|L> = -\eta^2 e^{2i\alpha}|L>$

Both the change of SAM and OAM is visible if we use odd no. of times the half wave-plates on LCP

$H(\phi)H(\phi+\alpha)H(\phi+\alpha+\beta)|L> = -i\eta^3 |R> e^{2i(\phi+\beta)}$ (18)

On the other hand with the action of even number of times the half wave plates on LCP gives change of OAM only without changing of SAM

$H(\phi)H(\phi+\alpha)H(\phi+\alpha)H(\phi+\alpha+\beta)|L> = \eta^4 |L> e^{2i(\alpha+\beta)}$ (19)

This shows the geometric phase developed from Half-wave plates by circularly polarized light visualize gain in orbital angular momentum of photon.